\documentclass[aps,showpacs,prl,superscriptaddress,reprint]{revtex4-1} 

\usepackage[colorlinks=true,linkcolor=blue,citecolor=blue,urlcolor=blue]{hyperref}

\usepackage{setspace} 
\usepackage{graphicx}
\usepackage{amsmath}
\usepackage{color}
\usepackage{amsmath}
\usepackage{amssymb}
\usepackage{verbatim}
\usepackage{latexsym}
\usepackage{enumerate} 
\usepackage{bm} 
\usepackage{ulem} 
\usepackage[caption=false]{subfig}
\usepackage{multirow}
\usepackage{booktabs}
\usepackage{lineno}

\begin{document}

\title{Tunable photostriction of halide perovskites through energy dependent photoexcitation}

\author{Bo Peng}
\affiliation{Theory of Condensed Matter Group, Cavendish Laboratory, University of Cambridge, J.\,J.\,Thomson Avenue, Cambridge CB3 0HE, United Kingdom}
\author{Daniel Bennett}
\affiliation{Theory of Condensed Matter Group, Cavendish Laboratory, University of Cambridge, J.\,J.\,Thomson Avenue, Cambridge CB3 0HE, United Kingdom}
\author{Ivona Bravi\'{c}}
\affiliation{Theory of Condensed Matter Group, Cavendish Laboratory, University of Cambridge, J.\,J.\,Thomson Avenue, Cambridge CB3 0HE, United Kingdom}
\author{Bartomeu Monserrat} \email{bm418@cam.ac.uk}
\affiliation{Theory of Condensed Matter Group, Cavendish Laboratory, University of Cambridge, J.\,J.\,Thomson Avenue, Cambridge CB3 0HE, United Kingdom}
\affiliation{Department of Materials Science and Metallurgy, University of Cambridge, 27 Charles Babbage Road, Cambridge CB3 0FS, United Kingdom}

\date{\today}

\begin{abstract}
Halide perovskites exhibit giant photostriction, that is, volume or shape changes upon illumination. However, the microscopic origin of this phenomenon remains unclear and there are experimental reports of both light-induced lattice expansion and contraction. In this paper we establish a general method, based on first-principles calculations and molecular orbital theory, which provides a microscopic picture of photostriction in insulators based on the orbital characters of their electronic bands near the Fermi level. For lead-halide perovskites, we find that different valence states have different bonding characters, leading to opposing strengthening or weakening of bonds depending on the photoexcitation energy. The overall trend is that light induces lattice contraction at low excitation energies, while giant lattice expansion occurs at high excitation energies, rationalizing experimental reports. 

\end{abstract}

\maketitle

Photostriction is the process of non-thermal deformation of materials under illumination. It has been well studied in ferroelectric materials such as SbSI \cite{Fridkin1967} and oxide perovskites \cite{Kundys2010,Wei2017a,Fu2020} due to promising applications in microactuators, sensors, and photonic devices \cite{Kundys2015}. These studies reveal that the driving mechanism behind photostriction can have different microscopic origins. For example, in polar materials the photoexcited carriers screen the polarization, which can change the internal electric field, inducing the converse piezoelectric effect \cite{Paillard2016,Paillard2017,Haleoot2017}. By contrast, in non-polar materials, the introduction of electron-hole pairs under illumination can cause changes in atomic bonds that lead to lattice deformations \cite{Kundys2015}. 

The recent discovery of giant photostriction in halide perovskites \cite{Zhou2016,Wei2017,Tsai2018} has been shown to boost the power conversion efficiency of perovskite-based solar cells \cite{Tsai2018}, to influence phase segregation \cite{Mao2021,Muscarella2020} and moisture degradation \cite{Lu2019}, and also provides insights into the observed light-induced intensity changes of photoluminescence \cite{Gottesman2015,DeQuilettes2016}. 
Additionally, photostriction could offer new pathways to use halide perovskites in optomechanical devices \cite{Wei2017,Zhou2016}.


However, the microscopic origin of photostriction in halide perovskites remains unclear. 
It is still debated whether the giant lattice expansion of up to 1.4\% under illumination can be induced by electronic excitation alone \cite{Tsai2018,Rolston2020,Tsai2020,Wei2017,Zhou2016a,Liu2019}. In addition, both light-induced lattice expansion and contraction have been observed in experiments: the giant lattice expansion is observed when the sample is illuminated by light sources such as a halogen lamp and a standard 1-sun source with excitation energies much greater than the band gap \cite{Tsai2018,Wei2017,Zhou2016a}, whereas lattice contraction 
occurs at an excitation energy of just 0.04 eV above the band gap \cite{Liu2019}. The observations of both lattice contraction and expansion, and the large magnitude of the volume changes, call for a better understanding of the microscopic mechanism behind photostriction in halide perovskites.


\begin{figure*}
\centering
\includegraphics[width=\linewidth]{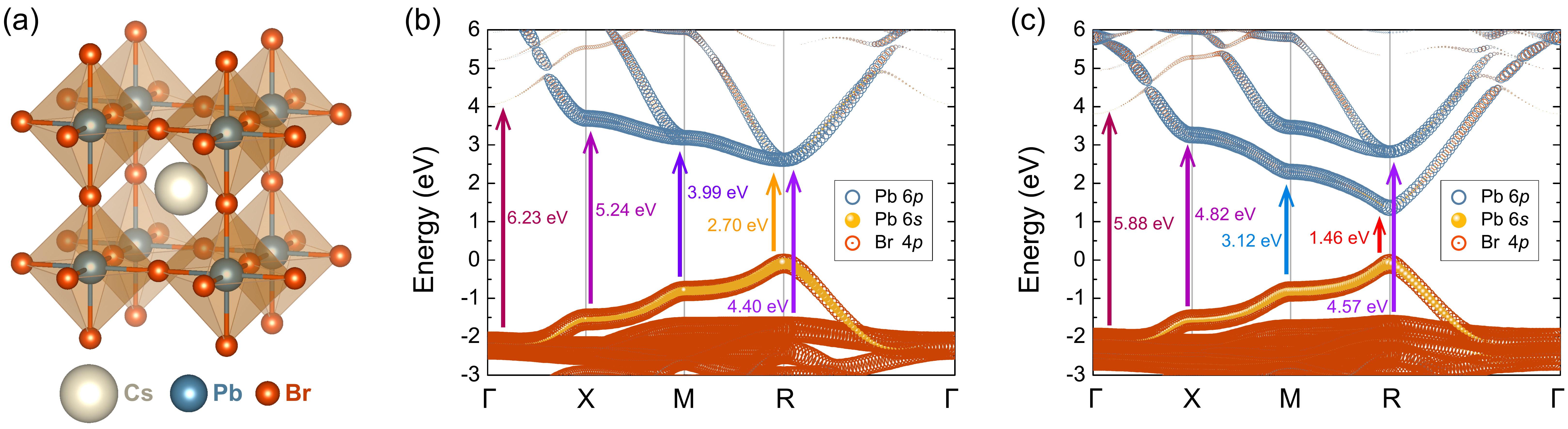}
\caption{(a) Crystal structure of CsPbBr$_3$. Electronic structure of CsPbBr$_3$ (b) without and (c) with spin-orbit coupling. The arrows indicate optical transition energies at different high symmetry points.}
\label{soc} 
\end{figure*}

In this work, we introduce a physically intuitive method to understand photostriction in semiconductors and insulators using a combination of electronic structure theory and molecular orbital theory, based on the nature of the orbital characters near the band edges. Motivated by contradictory experimental results in lead-halide perovskites, we study the effect of photoexcited carriers 
on the volume of cubic CsPbX$_3$ (X = Cl, Br, I), and illustrate that exciting electrons from valence bands with different orbital characters leads to tunable photostriction. 
When electrons are promoted from the strong antibonding states in the top valence band to the weaker antibonding conduction states, we obtain lattice contraction as a result of an overall weakened antibonding interaction. On the other hand, 
when electrons are promoted from the deeper, non-bonding valence bands to the antibonding conduction states, the antibonding interaction increases and leads to lattice expansion. 
Our findings pave the way to photocontrollable optoelectronic properties in lead-halide perovskites -- a desirable feature for technological applications. 
Overall, we propose that a detailed knowledge of the electronic structure and the band representations are the key ingredients to quantitatively understand photostriction in general insulators.



We use density functional theory (DFT) to study the ground state structural properties of cubic halide perovskites. We perform calculations using the {\sc vasp} code \cite{Kresse1996a} and cross-check them using the {\sc abinit} code \cite{Gonze2002} 
(see Section A in the Supplemental Material for computational details). At high temperatures, 
halide perovskites crystallize in a cubic ABX$_3$-type structure with space group $Pm\bar{3}m$, as shown in Figure\,\ref{soc}(a). The B atom sits at the center of an octahedral cage of halide X atoms, and the A atom occupies the cavity between neighboring octahedra. For halide perovskites, A is a monovalent cation like Cs or an organic molecule, B is a divalent metal cation like Sn or Pb, and X is a halogen anion like Cl, Br, or I. In this work, we focus on inorganic lead-halide perovskites, CsPbX$_3$, as representative examples of the whole family. Throughout the paper, we report the results for CsPbBr$_3$, and analogous results for $\mathrm{X = Cl}$ and I are discussed later. We focus on the high-temperature cubic phase and uniform strain for clarity, although the microscopic mechanism revealed by our analysis could be extended to also provide insights into the observed photoinduced phase transitions in halide perovskites \cite{Kirschner2019,Xue2019,Kim2019b}. Optimizing the crystal structure, we obtain a lattice constant of $5.859$ \AA\ for the unit cell of $Pm\bar{3}m$ CsPbBr$_3$ using the PBEsol functional \cite{Perdew2008} in {\sc vasp}, corresponding to a volume of 201.126 \AA$^3$. Although the photostriction results are independent of the choice of functional, we choose the PBEsol functional because it provides the best agreement with the measured value of $5.87$ \AA\ at $435$ K \cite{Rodova2003,Stoumpos2013} compared to other functionals (for details, see Section A in the Supplemental Material).

We calculate the electronic band structure both without [Figure\,\ref{soc}(b)] and with spin-orbit coupling (SOC) [Figure\,\ref{soc}(c)]. In light of the well-known feature of semilocal DFT to underestimate the band gap of semiconductors, we perform many-body perturbation theory calculations under the $G_0W_0$ approximation \cite{Shishkin2006,Shishkin2007} to obtain a more accurate band gap (see Section B in the Supplemental Material). We use a scissor shift calculated from the $G_0W_0$ band gap to correct the band structures obtained using DFT.






We study photostriction by using the $\Delta$ self-consistent field ($\Delta$SCF) method, where non-interacting electron-hole pairs are introduced within the Kohn-Sham formalism to simulate excited states \cite{Goerling1996,Jones1989,Hellman2004}. The accurate description of electronic excitations typically requires many-body perturbation theory methods \cite{Albrecht1998,Rohlfing1998,Shishkin2006,Shishkin2007,Fuchs2007,Sander2015} or time-dependent DFT \cite{Meng2008,Casida2012}. However, these methods are computationally expensive, and not always compatible with structural relaxations. Instead, the $\Delta$SCF method is a computationally simple approach that can readily describe structural relaxations in the presence of non-interacting electron-hole pairs, and has been found to provide good agreement with experiments in the context of photostriction \cite{Paillard2019,Marsili2013,Peng2020,Wang2020a}. We therefore perform $\Delta$SCF calculations using {\sc vasp} and also reproduce them using {\sc abinit} (see Section A in the Supplemental Material). We consider vertical excitations from/to different bands as specified below, and report
results according to the density $n$ of excited carriers resulting from these processes (for details, see Section C of the Supplemental Material).
The cubic lattice constant is then relaxed while keeping the electron occupations fixed. 



\begin{figure}
\centering
\includegraphics[width=0.4\textwidth]{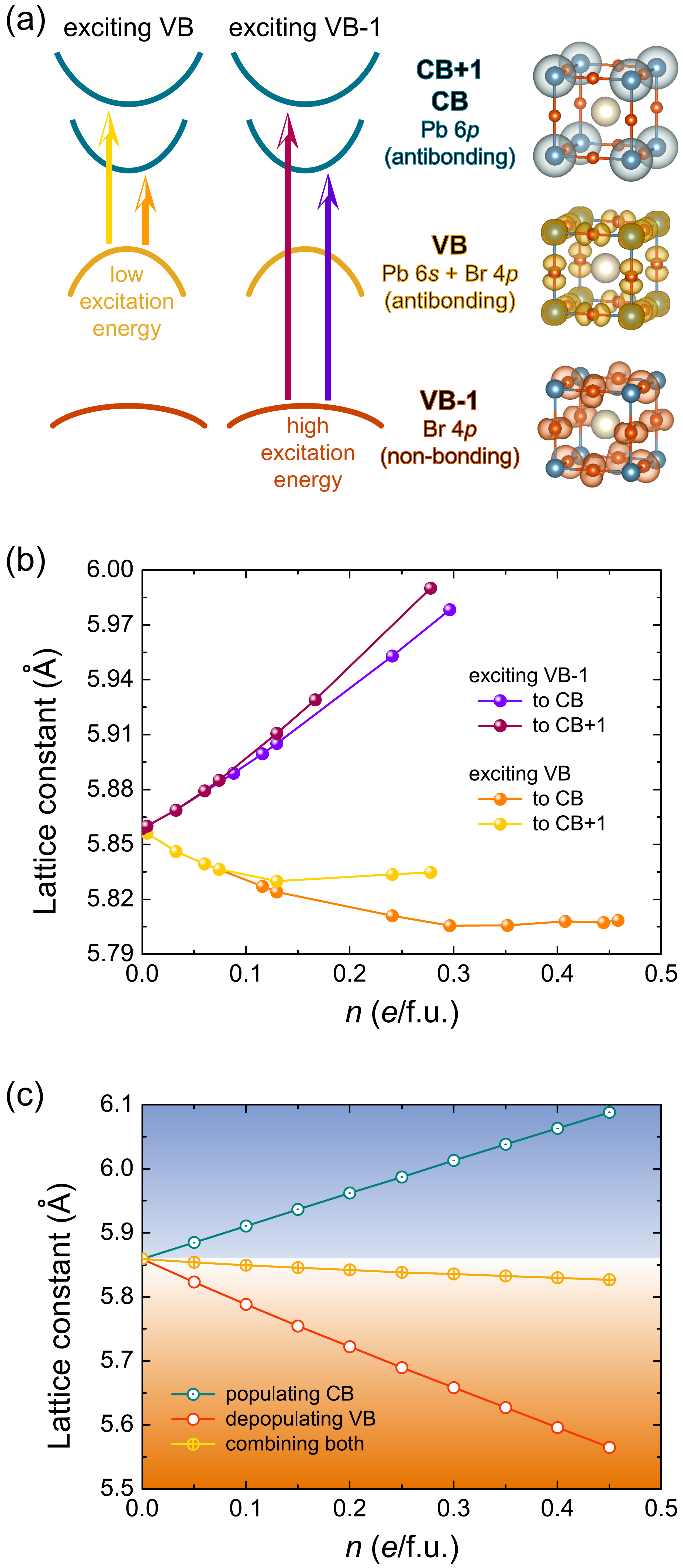}
\caption{(a) Selective excitation of the antibonding VB state and the non-bonding VB-1 state at the band gap of CsPbBr$_3$. Evolution of the lattice constant of CsPbBr$_3$ (b) with increasing photoexcited carrier density $n$ upon selective excitation, and (c) with increasing electron/hole doping concentration $n$ that populates/depopulates the CB/VB states.}
\label{schematic} 
\end{figure}

We first examine the orbital character of the electronic states involved in typical photoexcitation processes. We start from the case without SOC to demonstrate the role of orbital-selective electronic excitation, and include SOC later when discussing the overall photostrictive effects and the resulting optical properties upon photostriction. Without SOC, CsPbBr$_3$ has a direct band gap of $2.70$ eV at the R point $(0.5, 0.5, 0.5)$ from $G_0W_0$ calculations, as shown in Figure\,\ref{soc}(b). There are three types of orbitals 
involved in photoexcitation: (1) the conduction band minimum (CBM) at R is composed of triply degenerate (sixfold degenerate with spin), weak antibonding Pb 6$p$ orbitals, and the orbital character remains largely similar away from the R point; 
(2) for the top valence band, VB, the strongest antibonding interaction between the Pb 6$s$ orbitals and the Br 4$p$ orbitals is at the R point, and the antibonding interaction becomes weaker away from the R point as the contribution from the Pb 6$s$ orbitals decreases; (3) the valence band directly beneath the top one, VB-1, is less dispersive and has non-bonding character because it is mainly comprised of localized Br 4$p$ orbitals. 
The different bonding characters are illustrated in Figure\,\ref{schematic}(a) with the corresponding partial charge densities of the lowest conduction band CB, the top valence band VB, and the next valence band VB-1 at the R point.

To understand how bonding character affects photostrictive behavior, 
we first consider excitation processes between the VB and CB states. 
Based on the irreducible band representations (irreps), these excitations are allowed by symmetry.
The fundamental band gap of the system is located at the R point, which belongs to the point group $O_h$. 
The valence band maximum (VBM) arises from the antibonding hybridization of Pb 6$s$ orbitals and Br 4$p$ orbitals (irrep. R$_1^+$), and the CBM corresponds to an antibonding state of Pb 6$p$ orbitals (irrep. R$_4^-$). Selection rules demand that optical excitations in systems with inversion symmetry are allowed only when the two irreps of the electronic states have opposite parities, denoted with the descriptors even ($+$) and odd ($-$) in our notation. In this case, the VBM and the CBM have opposite parities, making the optical excitation dipole allowed (the full group theoretical analysis is available in Section D of the Supplemental Material). 
Despite both being antibonding states, the antibonding interaction in the VBM is stronger than that in the CBM because the former has a larger overlap between the Pb 6$s$ states and the Br 4$p$ states compared to the relatively isolated 6$p$ states around the Pb atoms in the CBM. As a result, the symmetry-allowed bright excitation from the VBM to the CBM redistributes an electron from a strong antibonding state to a weaker antibonding one, which consequently leads to a reduced antibonding interaction and to lattice contraction. 

To verify this picture computationally, we selectively excite electrons from the VB to the CB within the $\Delta$SCF approach and calculate the resulting relaxed volume, as shown by the orange curve in Figure\,\ref{schematic}(b). We start by only exciting electrons near the minimum band gap at the R point, leading to small photoexcited carrier densities ($n\lesssim 0.3$ $e$/f.u.). In this regime we observe lattice contraction.  
Exciting additional electrons from the VB to the CB at \textbf{k}-points that are increasingly away from the minimum band gap at R (with a corresponding increase in the photoexcited carrier density $n \gtrsim 0.3$ $e$/f.u.), 
involves the excitation from weakened antibonding states and leads to diminished lattice contraction. Continuing in this fashion, the antibonding interaction becomes so weak that exciting these states to the relatively stronger antibonding CB states results in lattice expansion [not shown in Figure\,\ref{schematic}(b), see Section E in the Supplemental Material for details]. The results of this computational experiment confirm the bonding picture for excitations between the VB and the CB.

To further explore the relative strength of the antibonding interactions in the VB and CB states, we calculate the evolution of the lattice constant upon doping, considering both the population of the CB states (electron doping) or the depopulation of the VB states (hole doping). As shown in Figure\,\ref{schematic}(c), when populating the CB states, the lattice expands, indicating an enhanced antibonding interaction. Similarly, the decreased lattice constant when depopulating the VB states demonstrates a weakened antibonding strength. The relative strength of the VB and CB states can be estimated from the slope of the corresponding red and blue curves in Figure\,\ref{schematic}(c). The decrease of lattice constant by depopulating the VB states is faster than the increase of lattice constant by populating the CB states, indicating that the antibonding strength of the VB is stronger than that of the CB. Therefore the average lattice constant decreases, as demonstrated by the yellow curve in Figure\,\ref{schematic}(c). This is consistent with our molecular orbital theory analysis in Section E of the Supplemental Material.


We next consider excitations from the VB-1 states. These non-bonding states do not contribute to the bonding-antibonding interaction, and therefore the depopulation of these states has negligible effects on the lattice constant. On the other hand, the optical transition populates the conduction bands of antibonding orbitals, increasing the interaction between the delocalized Pb 6$p$ orbitals and enhancing the repulsion between Pb atoms. Consequently, selective excitation from VB-1 to CB results in lattice expansion.

To verify this picture computationally, we selectively promote electrons from the VB-1 to the CB using the $\Delta$SCF method. We highlight that this computational experiment does not correspond to a physically realizable photoexcitation process because band overlaps and energy conservation may dictate that electrons from the VB should also be excited for a given photoexcitation energy, but it is a convenient model to isolate the role of the VB-1 band.
The results are shown by the violet curve in Figure\,\ref{schematic}(b), confirming the lattice expansion from the bonding picture.

For completeness, we also consider excitations from the VB and VB-1 bands to the CB+1 band in Figure\,\ref{schematic}(b). However, the character of the CB+1 band is similar to that of the CB band, so the results are largely independent of the conduction bands into which electrons are excited.

Overall, our combined analysis based on molecular orbital theory and first-principles calculations shows that the nature of photostriction is dominated by the changes in the valence states from which electrons are excited rather than the conduction states into which they are excited.


Having established the role of different bands in the photostriction response of CsPbBr$_3$ by (unphysically) exciting electrons from and to selected bands only, we next combine our results to rationalize experimental photostriction reports. We use the $\Delta$SCF method to create photoexcited configurations with fixed carrier densities. To achieve a given photoexcited carrier density, we promote the necessary number of electrons from the valence bands, starting from the VBM state and then promoting electrons with decreasing energy, and simultaneously populate the conduction bands starting from the CBM state and then occupying states with increasing energy. This assumes that, after photoexcitation, relaxation of electrons and holes to the band edges is a faster process than electron-hole recombination. The final configuration corresponds to a steady out-of-equilibrium state with electrons populating the lowest energy conduction bands and holes populating the highest energy valence bands.

To achieve this configuration experimentally, it is necessary to use photon energies at least up to the largest energy difference between the highest occupied conduction band and the lowest unoccupied valence band, and in the following we call this maximum energy difference the \textit{excitation energy}. We provide the relation between a given photoexcited carrier density and the corresponding excitation energy for CsPbBr$_3$ in Figure\,S6 of the Supplemental Material. To promote the required number of electrons to achieve a given photoexcited carrier density, it is also necessary to use appropriate fluences. The maximum reported carrier densities can be realized by an experimentally feasible fluence of about 10 mJ/cm$^2$ (see Section C of the Supplemental Material for details).

\begin{figure}
\centering
\includegraphics[width=0.4\textwidth]{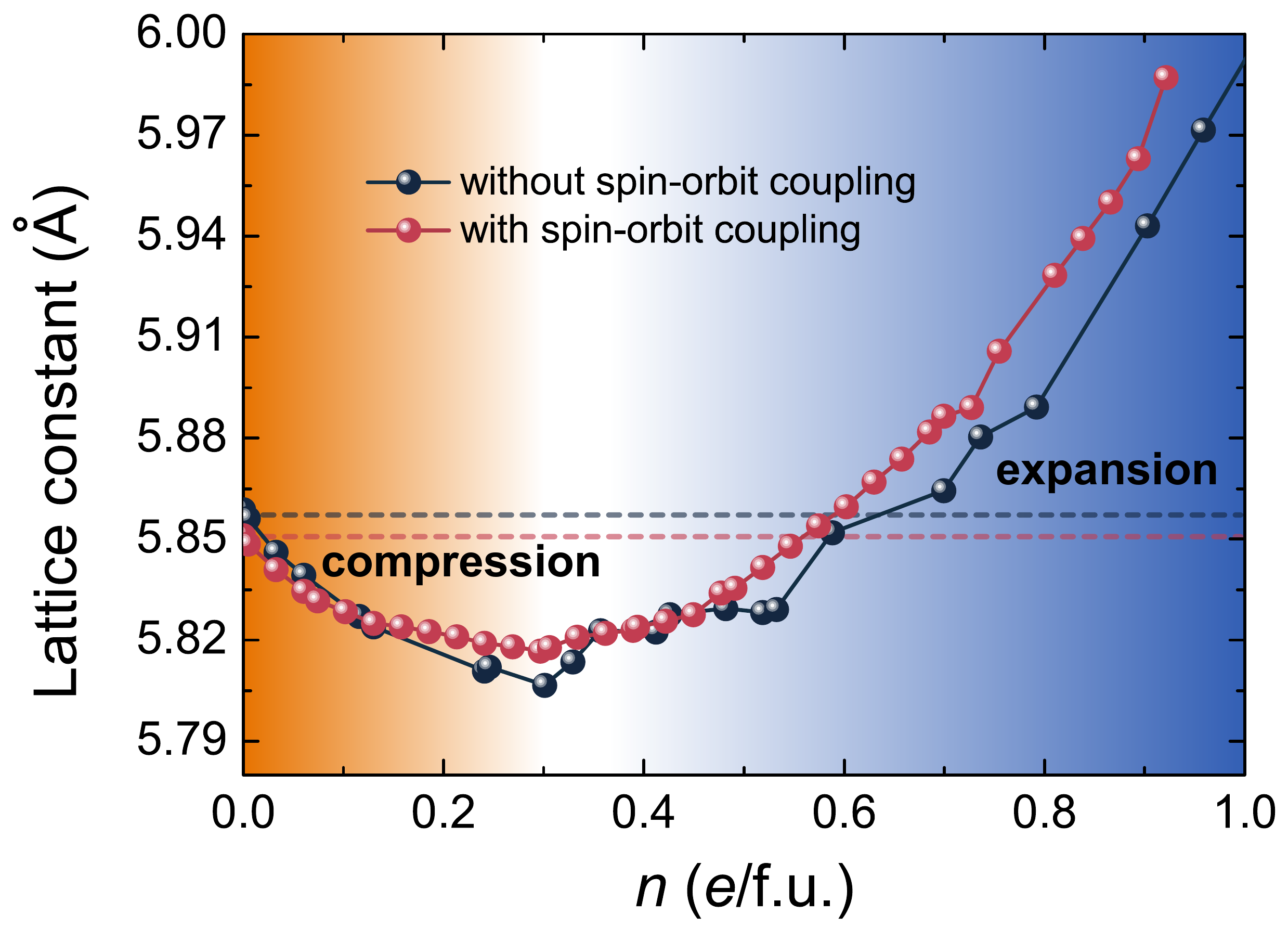}
\caption{Photostriction of CsPbBr$_3$ with and without spin-orbit coupling as a function of photoexcited carrier density. 
}
\label{total} 
\end{figure}

The volume of CsPbBr$_3$ as a function of photoexcited carrier density is depicted in Figure\,\ref{total}. For carrier densities below $n\lesssim 0.245$ $e/$f.u. (corresponding to excitation energies below 4.40 eV), only strong antibonding VB states are excited, leading to an overall lattice contraction. For higher carrier densities (excitation energies exceeding 4.40 eV), electrons from the VB-1 band start to be promoted, emptying the non-bonding VB-1 states and populating the antibonding conduction states. Therefore, the antibonding interaction is significantly enhanced, resulting in stronger repulsion between the Pb atoms which causes the lattice to expand. 
In this simple model, we obtain the maximum experimentally observed uniform strain of 1.4\% at a photoexcited carrier density $n \sim 0.8-0.9$ $e$/f.u. (corresponding to an excitation energy of $5.5-6.0$ eV). 

We then discuss the role of SOC in photostriction. The inclusion of SOC has a dramatic effect on the electronic structures of halide perovskites \cite{Katan2019}. As shown in Figure\,\ref{soc}(c), SOC splits the triply degenerate (sixfold degenerate when considering spin) CBM state at the R point into a doubly degenerate CB state and a fourfold degenerate CB+1 state \cite{Becker2018}. This spin-orbit splitting decreases the CB energy and increases the CB+1 energy, reducing the $G_0W_0$ band gap to 1.46 eV (see Section B in the Supplemental Material for a comparison of the calculated band gap with experiment).
Despite these SOC-driven changes to the band structure, the general microscopic mechanisms and resulting photostriction as a function of photoexcited carrier density remain largely unchanged, as shown in Figure\,\ref{total}. The robustness of the results with respect to the inclusion or exclusion of SOC arises because the band characters are similar between the two situations. However, this may not be true in other materials, especially when SOC induces band inversion between orbitals, and should be carefully checked.


The complete picture, summarized in Figure\,\ref{total}, shows that, at lower carrier densities (corresponding to low excitation energies) there is light-induced lattice contraction, while at larger carrier densities there is lattice expansion. 
These results allow us to rationalize the various experimental findings. It has been reported that by using halogen lamps or a standard 1-sun source with photoexcitation energies far above the band gap, lattice expansion of up to 1.4\% can take place 
\cite{Tsai2018,Wei2017,Zhou2016a}, but that a much smaller excitation energy of just 0.04 eV above the band gap 
induces compressive strain \cite{Liu2019}. 
Our study of the microscopic mechanism for photoexcitation in these materials explains the origin and conditions for both types of photostriction, providing a unified picture for the various experimental reports. We also note that photostriction should be considered when interpreting photoluminescence experiments (see Section F in the Supplemental Material for details).

Another contention point in the experimental study of photostriction is its microscopic origin. It is debated whether the lattice expansion in halide perovskites under illumination is induced by electronic excitation alone or by thermal expansion \cite{Tsai2018,Rolston2020,Tsai2020}, although we note that most experimental measurements have ruled out thermal expansion because the observed temperature change is only $0.3$\,K \cite{Tsai2018,Wei2017,Zhou2016a,Liu2019}. 
Our results also indicate that electronic excitation alone can induce giant photostriction of up to 2\% even without considering any thermal effects. 

\begin{figure}[h]
\centering
\includegraphics[width=0.4\textwidth]{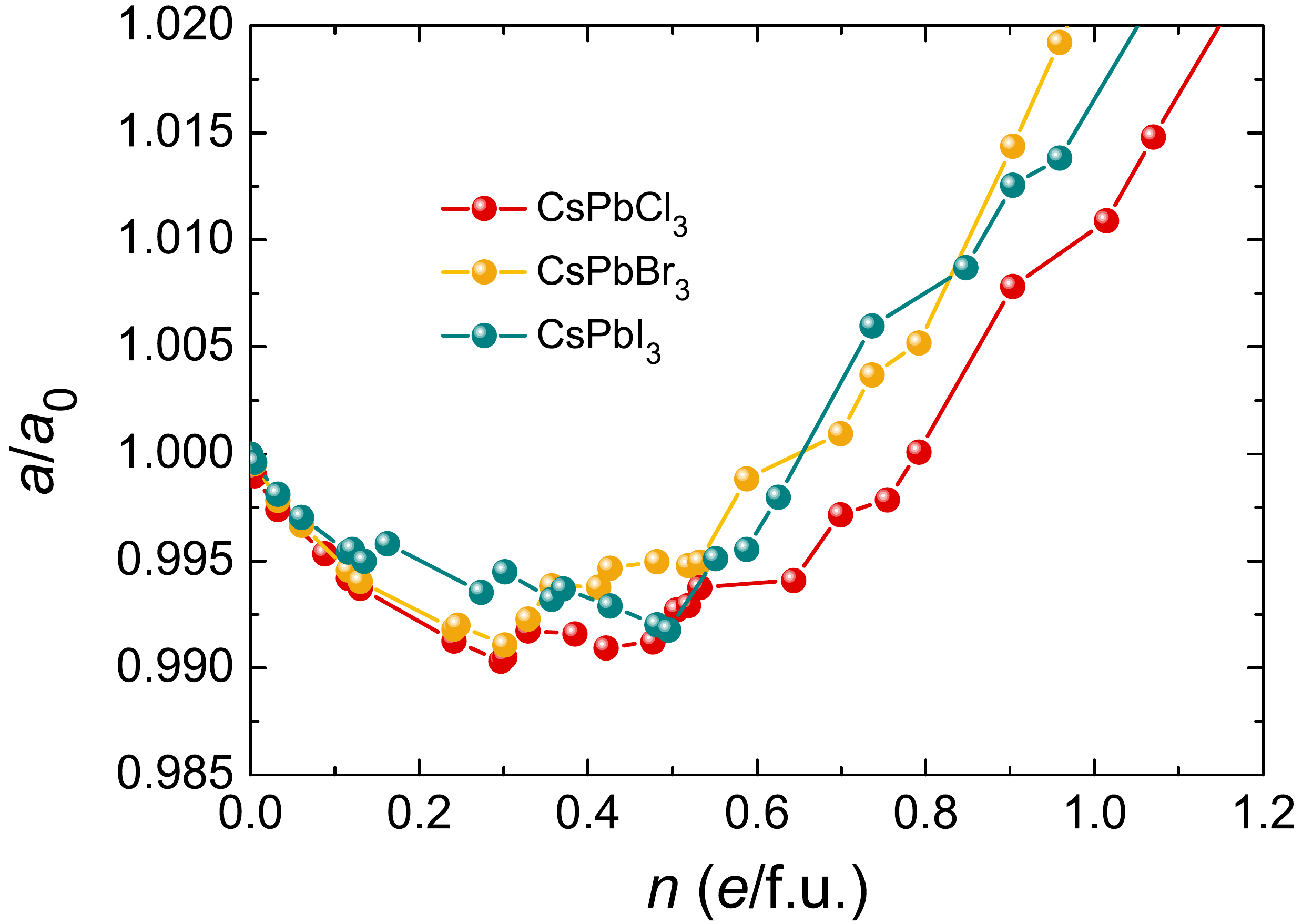}
\caption{
Change of the lattice constant $a$ compared to the lattice constant in the dark $a_0$ of CsPbCl$_3$, CsPbBr$_3$ and CsPbI$_3$ as a function of excited carrier density. All the results are obtained without $G_0W_0$ corrections and without SOC.}
\label{compounds} 
\end{figure}

The nonmonotonic trend of photostriction with increasing photoexcited carrier density is also observed in the other cesium lead-halide perovskites, CsPbCl$_3$ and CsPbI$_3$, as shown in Figure\,\ref{compounds}. This is unsurprising, since all lead-halide perovskites have similar orbital characters near the band edges. 

In summary, we provide a comprehensive microscopic description of photostriction in lead-halide perovskites starting from a simple consideration of the orbital characters of their band structure. When exciting from strong antibonding states in the top valence band to the weaker antibonding conduction states, lattice contraction results from a weakened antibonding interaction. In contrast, for larger excitation energies that promote electrons from the deeper, non-bonding valence bands to the antibonding conduction states, the resulting enhanced antibonding interaction leads to lattice expansion. The overall effect is that light induces lattice contraction at low excitation energies, while lattice expansion occurs at higher excitation energies. Our findings help to rationalize the experimental observation that lattice expansion of up to 1.4\% can be induced by light of energy far above the band gap \cite{Tsai2018,Wei2017,Zhou2016a}, while excitation energies close to the gap induce compressive strain that contracts the lattice \cite{Liu2019}. The ability to control photostriction via the excitation energy may offer new opportunities for solar cells, light-emitting diodes, and memory devices with ultrafast control of their optical properties. More generally, our approach illustrates the important role that the orbital characters of the electron bands can play in photoexcited phenomena. The combination of molecular orbital theory and first-principles calculations should be generally applicable to understand photoinduced phenomena in other systems. 


\begin{acknowledgements}

We acknowledge helpful discussions with Dr Gaozhong Wang (Trinity College Dublin) on the estimation of photoexcited carrier density. B.P., I.B. and B.M. acknowledge support from the Winton Programme for the Physics of Sustainability. D.B. acknowledges support from the EPSRC Centre for Doctoral Training in Computational Methods for Materials Science under grant number EP/L015552/1. B.M. also acknowledges support from a UKRI Future Leaders Fellowship (Grant No. MR/V023926/1) and from the Gianna Angelopoulos Programme for Science, Technology and Innovation. The calculations were performed using resources provided by the Cambridge Tier-2 system, operated by the University of Cambridge Research Computing Service (www.hpc.cam.ac.uk) and funded by EPSRC Tier-2 capital grant EP/P020259/1, as well as with computational support from the U.K. Materials and Molecular Modelling Hub, which is partially funded by EPSRC (EP/P020194), for which access is obtained via the UKCP consortium and funded by EPSRC grant ref. EP/P022561/1.

\end{acknowledgements}


%

\end{document}